\documentclass[showpacs,preprintnumbers,amsmath,amssymb,showkeys,aps,prd,a4paper]{revtex4}

\usepackage[dvips]{graphicx}
\usepackage{dcolumn}
\usepackage{bm}
\usepackage{epsfig}
\usepackage{amsmath}
\usepackage{amsfonts}
\usepackage{amssymb,amscd}
\usepackage[latin1]{inputenc}
\usepackage{units}
\usepackage{hyperref}
\usepackage{braket}

\begin{document}

\begin{flushright}
LU TP 15-50\\
November 2015
\vskip1cm
\end{flushright}

\title{Probing the gluon density of the proton in the exclusive photoproduction of vector mesons at the LHC: A phenomenological analysis}
\author{V. P. Gon\c{c}alves$^{1,2}$, L. A. S. Martins$^{2}$ and W. K. Sauter$^{2}$}

\affiliation{ $^{1}$ Department of Astronomy and Theoretical Physics, Lund University, 223-62 Lund, Sweden.}
\affiliation{$^{2}$ Instituto de F\'{\i}sica e Matem\'atica,  Universidade
Federal de Pelotas, 
Caixa Postal 354, CEP 96010-900, Pelotas, RS, Brazil}

\date{\today}

\begin{abstract}
The current uncertainty on the gluon density extracted from the global parton analysis is large in the kinematical range of small values of the Bjorken - $x$ variable and low values of the hard scale $Q^2$. An alternative to reduces this uncertainty is the analysis of the  exclusive vector meson photoproduction in photon - hadron and hadron - hadron collisions. This process offers a unique opportunity to constrain the gluon density of the proton, since its cross section is proportional to the gluon density squared. In this paper we consider current parametrizations for the gluon distribution  and estimate the exclusive vector meson photoproduction cross section at HERA and LHC using the leading logarithmic formalism. We perform a fit of the normalization of the $\gamma h$ cross section and the value of the hard scale for  the process and  demonstrate that the current LHCb experimental data are better described by models that assume a slow increasing of the gluon distribution at small - $x$ and low $Q^2$. 
\end{abstract}

\keywords{Vector meson production, exclusive processes, parton distributions}
\pacs{12.38.Bx, 13.60.Le}

\maketitle

\begin{figure}[t]
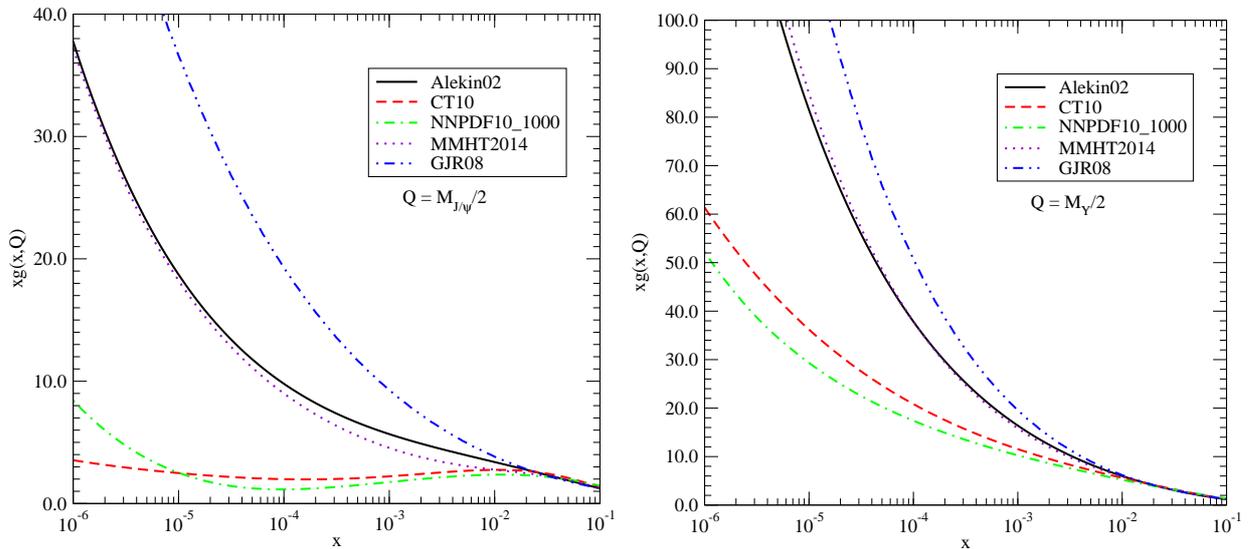

\begin{center}
\begin{tabular}{lr}
 \includegraphics*[width=0.45\columnwidth]{gluon_pdf.eps} & \includegraphics*[width=0.45\columnwidth]{gluon_pdf_upsilon.eps}
\end{tabular}
\end{center}
\caption{\label{fig:xgqx} Comparison between the gluon distributions obtained by different groups \cite{Alekhin:2002fv,Lai:2010vv,Ball:2008by,Harland-Lang:2014zoa,Gluck:2007ck} in the global analysis of the experimental data. We present the results  for two different values of the hard scale:  $Q = M_{J/\psi}/2$ (left panel) and  $M_{\Upsilon}/2$ (right panel).}
\end{figure}

One of the basic ingredients to estimate the hadronic cross sections are the parton distributions functions (PDFs). Theoretically, at large energies the hadrons are dominated by gluons, with its behaviour at small-$x$ being determined by the QCD dynamics at high parton densities. Consequently, a precise determination of the gluon distribution is fundamental to probe a possible transition between the linear and non linear regimes of the QCD dynamics \cite{hdqcd}. Experimentally, our understanding about the partonic structure of the proton has been significantly improved by the results obtained in $ep$ collisions at HERA, which have obtained very precise data in a broad range in photon virtualities  $Q^2$ and Bjorken - $x$ values, imposing the tightest constraints on the existing PDFs (For  recent reviews see, e.g. Refs. \cite{forte,hera,Abramowicz:2015mha}). However, the behaviour of the gluon distribution at small- $x$ is still poorly known as can be observed in Fig. \ref{fig:xgqx}, where we compare the predictions for the gluon distribution  obtained by different groups 
 \cite{Alekhin:2002fv,Lai:2010vv,Ball:2008by,Harland-Lang:2014zoa,Gluck:2007ck}  that perform the global analysis of the existing experimental data using the DGLAP evolution equations \cite{dglap} in order to determine the parton distributions. Consequently, additional measurements are necessary to pin down the gluon distribution. An alternative is the analysis of the experimental results for the heavy quark production at forward rapidities in $pp$ collisions at the LHC energies. Recent results \cite{prosa,rojo,mangano} have analysed the impact of the LHCb data in the determination of the gluon distribution. Another promising observable is the cross section for the diffractive production of vector mesons, which in the leading logarithmic approximation  is proportional to the gluon density squared \cite{ryskin,brod}. This process was analysed  at HERA and is currently been studied in photon - induced interactions at hadronic colliders.  In recent  years a series of experimental results at RHIC \cite{star,phenix}, Tevatron \cite{cdf} and LHC \cite{alice, alice2,alice3,lhcb,lhcb2,lhcb_ups,cms1,cms2,cms3}  demonstrated that the study of photon - induced interactions in hadronic colliders is feasible and can be used to probe e.g. the nuclear gluon  distribution \cite{gluon,strikman,gluon2,gluon3,Guzey,Jones1,Jones2}, the dynamics of the strong interactions \cite{vicmag_mesons1,outros_vicmag_mesons,vicmag_update,motyka_watt,Lappi,griep,bruno1,bruno2}, the Odderon \cite{vicodd1,vicodd2}, the mechanism  of quarkonium production \cite{Schafer,mairon1,mairon2,cisek,bruno1,bruno2} and the photon flux of the proton \cite{vicgus1,vicgus2}. It has stimulated the improvement of the theoretical description of these processes as well as the proposal of new forward detectors to be installed in the LHC \cite{ctpps}.

The basic idea in the photon-induced processes is that
 a ultra relativistic charged hadron (proton or nuclei)
 give rise to strong electromagnetic fields, such that the photon stemming from the electromagnetic field
of one of the two colliding hadrons can interact with one photon of
the other hadron (photon - photon process) or can interact directly with the other hadron (photon - hadron
process) \cite{upc,epa}. In photon-hadron processes the total cross section  can be factorized in
terms of the equivalent flux of photons into the hadron projectile and the  photon-hadron production cross section, with the corresponding rapidity distribution being a direct probe of the photon - hadron cross section for a given energy. In the particular case of vector meson photoproduction in $pp$ collisions, the experimental data for a given rapidity $y$ gives access to the behaviour of the  gluon distribution of the proton for   $x = M_V/\sqrt{s} \exp(-y)$, where $M_V$ is the mass of the vector meson and $\sqrt{s}$ is the center - of - mass energy of the hadron - hadron collision. 
This property was the main motivation for the proposition presented  in Ref. \cite{gluon}, which was improved by  several authors in the last fourteen years \cite{strikman,gluon2,gluon3,Guzey,Jones1,Jones2}. Our goal in this paper is to complement these previous studies by performing a phenomenological analysis of the exclusive vector meson photoproduction which can illuminate several aspects of the formalism and about the current parametrizations for the gluon distribution obtained in the global parton analysis. Basically, we consider different models for $xg$ and estimate the cross section for the photoproduction of vector mesons in $\gamma p$ interactions using the leading logarithmic formalism \cite{ryskin,brod}. We demonstrate that assuming the usual value of the factorization scale $\bar{Q} = M_V/2$, the HERA and LHCb data for the $J/\Psi$ photoproduction are not described by these models. Taking into account that this formalism  have been derived at leading order, which implies a uncertainty in the choice of $\bar{Q}$, we determine its value by fitting the $\gamma h$ data. Using these best fit parameters we estimate the rapidity distributions for the $J/\Psi$, $\Psi(2S)$  and $\Upsilon$ production in $pp$ collisions at the LHC and compare with the recent LHCb data \cite{lhcb,lhcb2,lhcb_ups}.  Our results indicate that the current experimental data for the exclusive vector meson production in $pp$ collisions can only be described by models which predict a slow increasing  of the gluon distribution at small - $x$.

\begin{figure}[t] 
\begin{center}
\includegraphics*[width=0.5\columnwidth]{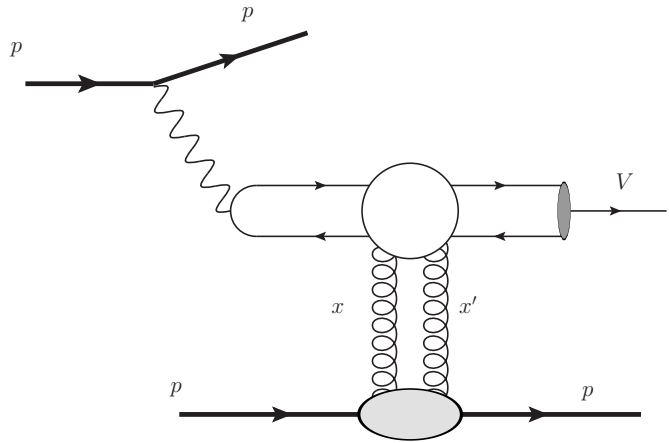}
\end{center}
\caption{Exclusive vector meson photoproduction in $pp$ collisions.} \label{fig:dia}
\end{figure} 

Lets initially present a brief review of the formalism for the calculation of the vector meson production in $pp$ collisions. The process is represented in Fig. \ref{fig:dia}, with its rapidity distribution being given by
\begin{equation} \label{eq:dsdy}
 \frac{d\sigma}{dy}(h_1 h_2 \rightarrow h_1 \otimes V \otimes h_2) = \mathcal{S}^2(W_+)\left[\omega_+\frac{dN_{\gamma/h_1}(\omega_+)}{d\omega_+}\right]\sigma_{\gamma h_2\rightarrow V h_2}(y) + \mathcal{S}^2(W_-)\left[\omega_-\frac{dN_{\gamma/h_2}(\omega_-)}{d\omega_-}\right]\sigma_{\gamma h_1 \rightarrow V h_1}(-y) 
\end{equation}
where we have taken into account that the two incident protons can be the source of the photon and $\otimes$ characterizes the presence of a rapidity gap in the final state. The photon - hadron center of mass energies squared $W^2_\pm$ and the photon energies $\omega_\pm$ are given by
\begin{equation}
 W^2_\pm = e^{\pm |y|}\sqrt{s}M_V,\quad \omega_\pm = \frac{M_V}{2}e^{\pm |y|}, 
\end{equation}
where $\sqrt{s_{h_1h_2}}$ is the hadron - hadron center - of - mass energy, $M_V$ is the mass of the vector mass and $y$ its rapidity.
Moreover, the factors $\mathcal{S}^2(W_\pm)$ characterizes absorptive corrections which can destroy the rapidity gaps generated in exclusive processes \cite{Jones1,Jones2}. The elastic photon flux for the proton can be expressed by \cite{dz}
\begin{eqnarray}
\frac{dN_{\gamma/p}(\omega)}{d\omega} =  \frac{\alpha_{\mathrm{em}}}{2 \pi\, \omega} \left[ 1 + \left(1 -
\frac{2\,\omega}{\sqrt{s_{NN}}}\right)^2 \right] 
\left( \ln{\Omega} - \frac{11}{6} + \frac{3}{\Omega}  - \frac{3}{2 \,\Omega^2} + \frac{1}{3 \,\Omega^3} \right) \,,
\label{eq:photon_spectrum}
\end{eqnarray}
with the notation $\Omega = 1 + [\,(0.71 \,\mathrm{GeV}^2)/Q_{\mathrm{min}}^2\,]$ and 
$Q_{\mathrm{min}}^2= \omega^2/[\,\gamma_L^2 \,(1-2\,\omega /\sqrt{s_{NN}})\,] \approx 
(\omega/
\gamma_L)^2$. The main input for the calculation of the rapidity distribution [ Eq. (\ref{eq:dsdy})] is the photon - hadron cross section for the vector meson production $\sigma_{\gamma h \rightarrow V h}$.
In the leading logarithmic approximation, the cross section for the  vector meson production  off any hadronic target, including a nucleus, at small-$x$ and for a sufficiently hard scale, is proportional to the square of the gluon parton density of the target.
 To lowest order, the $\gamma h \rightarrow V  h$ $(h = p, \,A)$
amplitude can be factorized into the product of the $\gamma
\rightarrow q \overline{q}$ transition $(q = c, \,b)$, the scattering of the
$q\overline{q}$ system on the target via (colorless) two-gluon
exchange, and finally the formation of the quarkonium from the
outgoing $q\overline{q}$ pair.  The heavy meson mass $M_{V}$
ensures that perturbative QCD can be applied to photoproduction.
The calculation was
performed some years ago to leading logarithmic ($\log(\overline{Q}^2)$) approximation,
assuming the produced vector meson quarkonium system to be
nonrelativistic \cite{ryskin,brod} and improved in distinct aspects
\cite{levin,fran}.
 Assuming a non-relativistic wave function for the vector meson one have that the $t=0$
differential cross section of photoproduction of heavy vector
mesons in leading order approximation  is given by \cite{ryskin,brod}
\begin{eqnarray}
 \left.\frac{d\sigma^{\gamma h \rightarrow Vh}}{dt}\right|_{t=0} = {\cal{N}} \frac{\pi^3 \Gamma_{e^+ e^-}M_V^3}{48 \alpha_\mathrm{em}}\left[ \frac{\alpha_s(\bar{Q}^2)}{\bar{Q}^4} xg_h(x, \bar{Q}^2) \right]^2
\,\,, \label{sigela}
\end{eqnarray}
where $xg_h$ is the target gluon distribution and  $x =
4\overline{Q}^2/W^2$, with $W$ the $\gamma h$ center of mass energy,
$\overline{Q}^2$ the characteristic hard scale of the processes and ${\cal{N}} = 1$ at leading order (see below). Moreover, $\Gamma_{ee}$ is the
leptonic decay width of the vector meson. In Refs. \cite{levin,fran,teubner,martin_ups,ivanov_nlo} the authors have estimated the
relativistic corrections [${\cal{O}}(4\%)$] , the real part contribution of the production amplitude [${\cal{O}}(15\%)$], the skewness effect of off-diagonal partons [${\cal{O}}(20\%)$] and next-to-leading order corrections [${\cal{O}}(40\%)$]
 to the LO exclusive heavy vector meson production, given by Eq. (\ref{sigela}). 
It is important to emphasize that magnitude of these corrections is still a subject of discussions \cite{Ivanov,Ivanov_gdp,Jones3}. In particular, the value of the hard scale $\bar{Q}$ is not fixed reliably at leading order. Such corrections have direct impact in the normalization of the cross section and in its energy dependence.  In what follows we take into account of the contributions associated to real part of the scattering amplitude and to the skewness effect, which is equivalent to multiply the Eq. (\ref{sigela}) by a factor $(1 + \beta) R_g^2$, where 
\begin{equation}
 \beta = \tan \frac{\pi \lambda}{2},\quad 
 R_g = \frac{2^{2\lambda + 3}}{\sqrt{\pi}} \frac{\Gamma(\lambda + 5/2)}{\Gamma(\lambda + 4)}
\end{equation}
with
\begin{equation}
\lambda = \frac{\partial \ln[xg(x,\bar{Q}^2)]}{\partial \ln 1/x}\,\,.
\end{equation}
 In order to estimate the total cross section we assume  an exponential parametrization for the small $|t|$ behaviour of the amplitude, which implies 
\begin{equation}
 \sigma_{\gamma h \rightarrow Vh} = \frac{1}{b_V} \left.\frac{d\sigma^{\gamma h \rightarrow Vh}}{dt}\right|_{t=0}
\end{equation}
with  $b_V$ being given by the following parametrization \cite{Jones1}
\[
 b_V(W) = 4.9 + 0.24 \ln (W/\unit[90]{GeV}),
\]
which is compatible with the HERA data. Moreover, we consider the values of mass and electronic decay widths of the vector mesons as given in Ref. \cite{Agashe:2014kda}.
Finally, in our calculations of the rapidity distribution we will assume that the absorptive corrections $\mathcal{S}^2(W_\pm)$ are given by the gap survival probability computed in \cite{Jones1,Jones2} using the model proposed in Ref. \cite{Khoze:2013dha}.

\begin{figure}[t] 
\begin{center}
\includegraphics*[width=0.9\columnwidth]{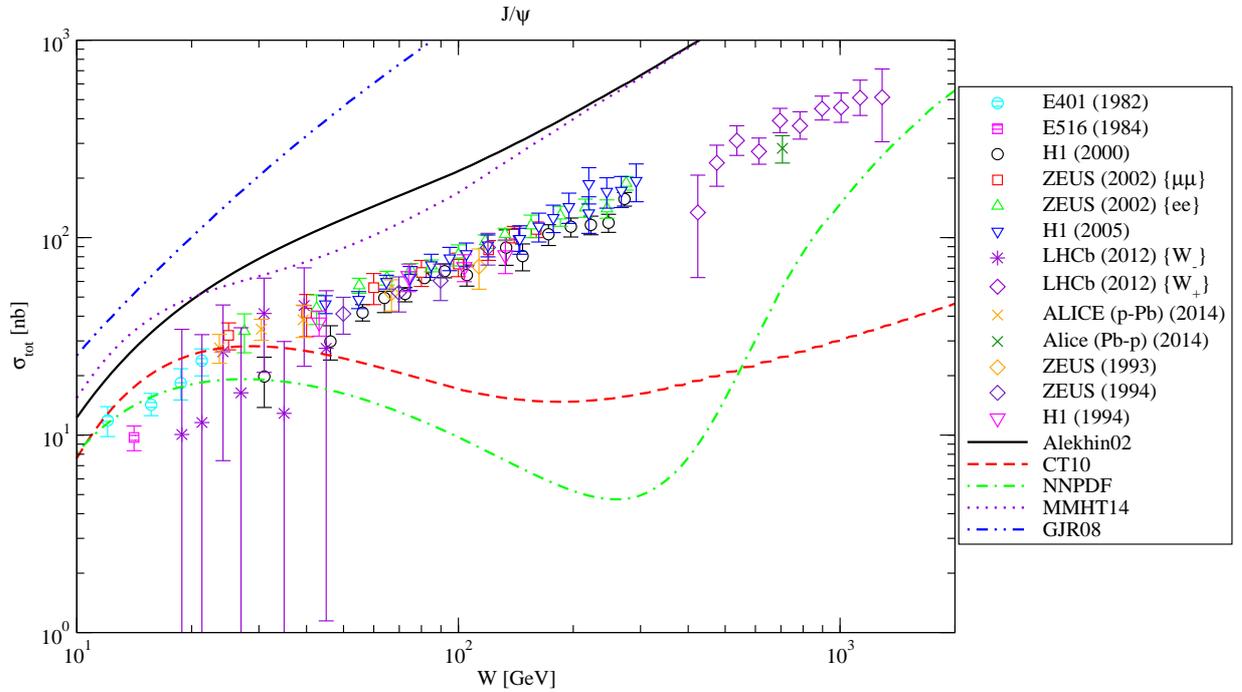}
\end{center}
\caption{Energy dependence of the  $J/\Psi$  photoproduction cross section obtained assuming that $\bar{Q} = M_{J/\Psi}/2$. Data from Refs.  
\cite{alice3,lhcb,lhcb2,H1_jpsi,h1A,h1B,ZEUS_jpsi}. } \label{fig:semesc}
\end{figure}

In Fig. \ref{fig:xgqx} we present the predictions for the gluon distribution obtained at leading order by the different groups \cite{Alekhin:2002fv,Lai:2010vv,Ball:2008by,Harland-Lang:2014zoa,Gluck:2007ck} that perform the global parton analysis of the experimental data to extract the parton distributions. We show the results for two different values for the hard scale: $Q = M_{J/\Psi}/2$ (left panel) and $Q = M_{\Upsilon}/2$ (right panel). These values are usually assumed as being $\bar{Q}$ in the calculations of the exclusive $J/\Psi$ and $\Upsilon$ photoproduction cross sections, respectively.  We have that the distinct predictions differ significantly, mainly at small - $x$ and low $Q$, which demonstrate that the global analysis do not reliably determine the gluon in this kinematical range. Basically, these distinct behaviours of the gluon distribution for low values of the hard scale are directly connected with the different assumptions for the initial conditions of the DGLAP equations considered by the distinct groups. With the increasing of the hard scale, the predictions becomes less dependent of these assumptions, and the distinct predictions becomes similar at very large $Q$. These results  demonstrate the importance of probes of the gluon distribution at small - $x$ and low $Q$.  
In Fig. \ref{fig:semesc} we present the resulting predictions for the energy dependence of the exclusive $J/\Psi$ photoproduction cross section obtained from Eq. (\ref{sigela}) assuming $\bar{Q} = M_{J/\Psi}/2$. The experimental data   from Refs.  
\cite{alice3,lhcb,lhcb2,H1_jpsi,h1A,h1B,ZEUS_jpsi} are presented for comparison. We obtain that the different models for the gluon distribution are not able to describe the normalization and/or the energy dependence of the data. The differences present in Fig.  \ref{fig:xgqx} are amplified in the 
exclusive $J/\Psi$ photoproduction cross section due to its quadratic dependence on $xg$. In particular, the distinct $x$ - behaviour predicted by the NNPDF and CT10 parametrizations for   
$Q = M_{J/\Psi}/2$, with $xg$ decreasing in the range  $10^{-4} \le x \le 10^{-2}$ before to increase for $x < 10^{-4}$, implies the anomalous behaviour in $\sigma_{\gamma p \rightarrow J/\Psi p}$ observed in Fig.  \ref{fig:semesc}.

A possible interpretation of the results presented in Fig. \ref{fig:semesc} is that the different models fail to describe the data since  our predictions for 
$\sigma_{\gamma p \rightarrow J/\Psi p}$ have been obtained  at leading order, which can be strongly affected by theoretical uncertainties associated to higher order corrections \cite{Ivanov,Ivanov_gdp}. It is generally believed that these higher order corrections should be important, but a full calculation still remains a challenge. In order to analyse the possible impacts of the higher order corrections in our phenomenological analysis, in what follows we will assume that the general behaviour of the cross section, Eq. (\ref{sigela}), will remain unchanged after the inclusion of the higher order corrections and that they can be effectively incorporated by taking the values of the normalization and the hard scale from a fit to the $\gamma h$ data. In other words, we will assume that the  main effect of the higher order corrections will be the  modification of the normalization of the cross section and that they can be taken into account by an appropriate choice of the hard scale (For a similar approach see \cite{Guzey}).  Basically we will assume $\bar{Q} = \xi M_V/2$ and will take $\xi$ and ${\cal{N}}$ in Eq. (\ref{sigela}) as free parameters to be determined by fitting the experimental data for the $\sigma_{\gamma p \rightarrow V p}$ cross section. We consider
the experimental data from Fermilab, HERA and LHC for  the $J/\Psi$, $\Psi(2S)$ and $\Upsilon$ production and performed the minimization of $\chi^2/\mathrm{d.o.f.}$ in order to determine  $\xi$ and ${\cal{N}}$. These new parameters are then used to calculate the rapidity distribution for the vector meson photoproduction in $pp$ collisions.   

\begin{figure}
\begin{center}
\begin{tabular}{cc}
 \includegraphics[scale=0.75]{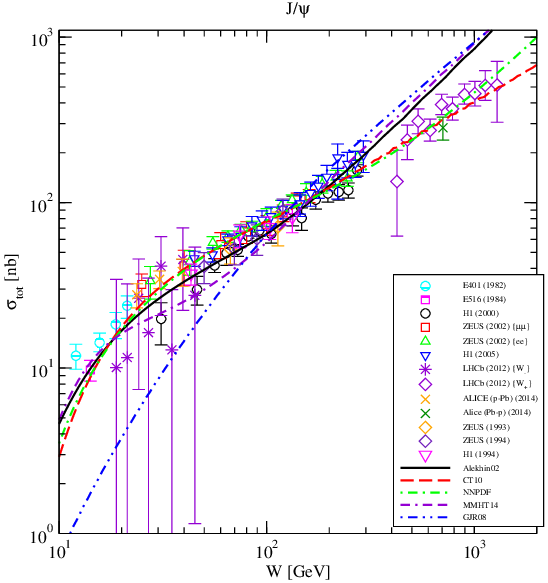} & \includegraphics[scale=0.85]{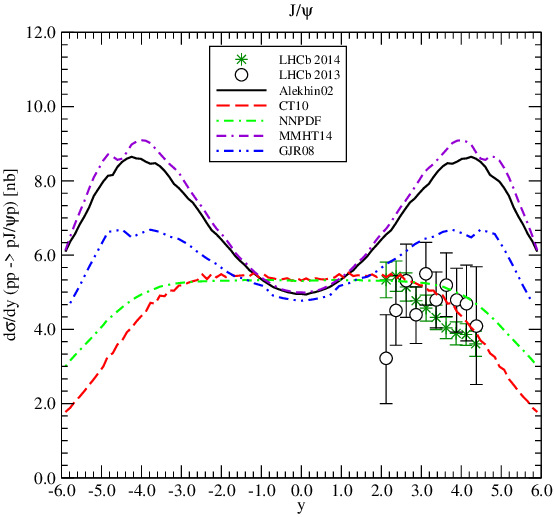}
\end{tabular}
\end{center}
\caption{\label{fig:jpsi} Left panel: Energy dependence of the exclusive  $J/\psi$ photoproduction cross section obtained using the parameters described in the Table \ref{tab:param}. Right panel: Predictions for the rapidity distribution for the exclusive $J/\psi$ photoproduction in $pp$ collisions at $\sqrt{s} = 7$ TeV. Data from Refs.  
\cite{alice3,lhcb,lhcb2,H1_jpsi,h1A,h1B,ZEUS_jpsi}.}
\end{figure}

\begin{figure}
\begin{center}
\begin{tabular}{cc}
 \includegraphics[scale=0.35]{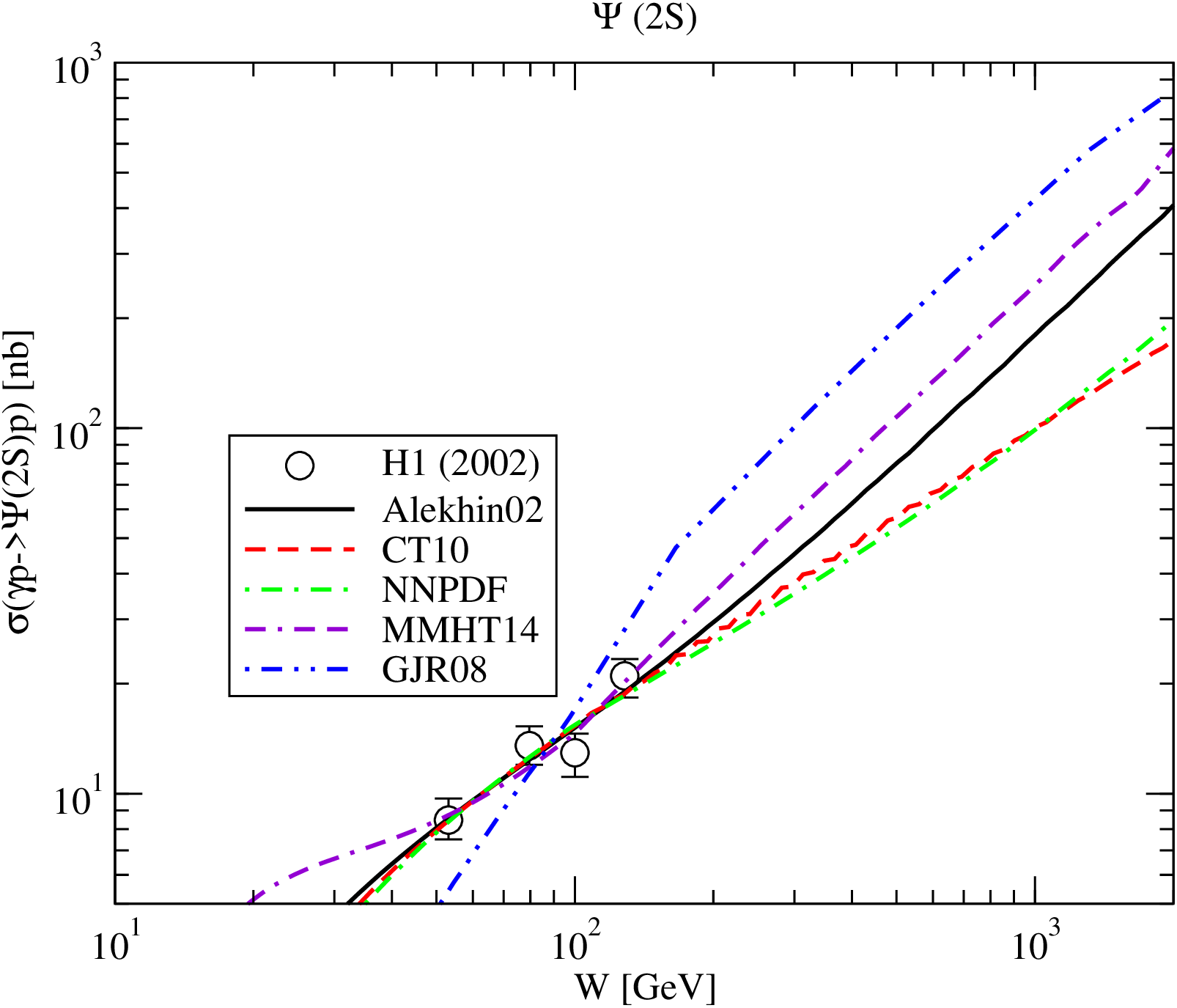} & \includegraphics[scale=0.45]{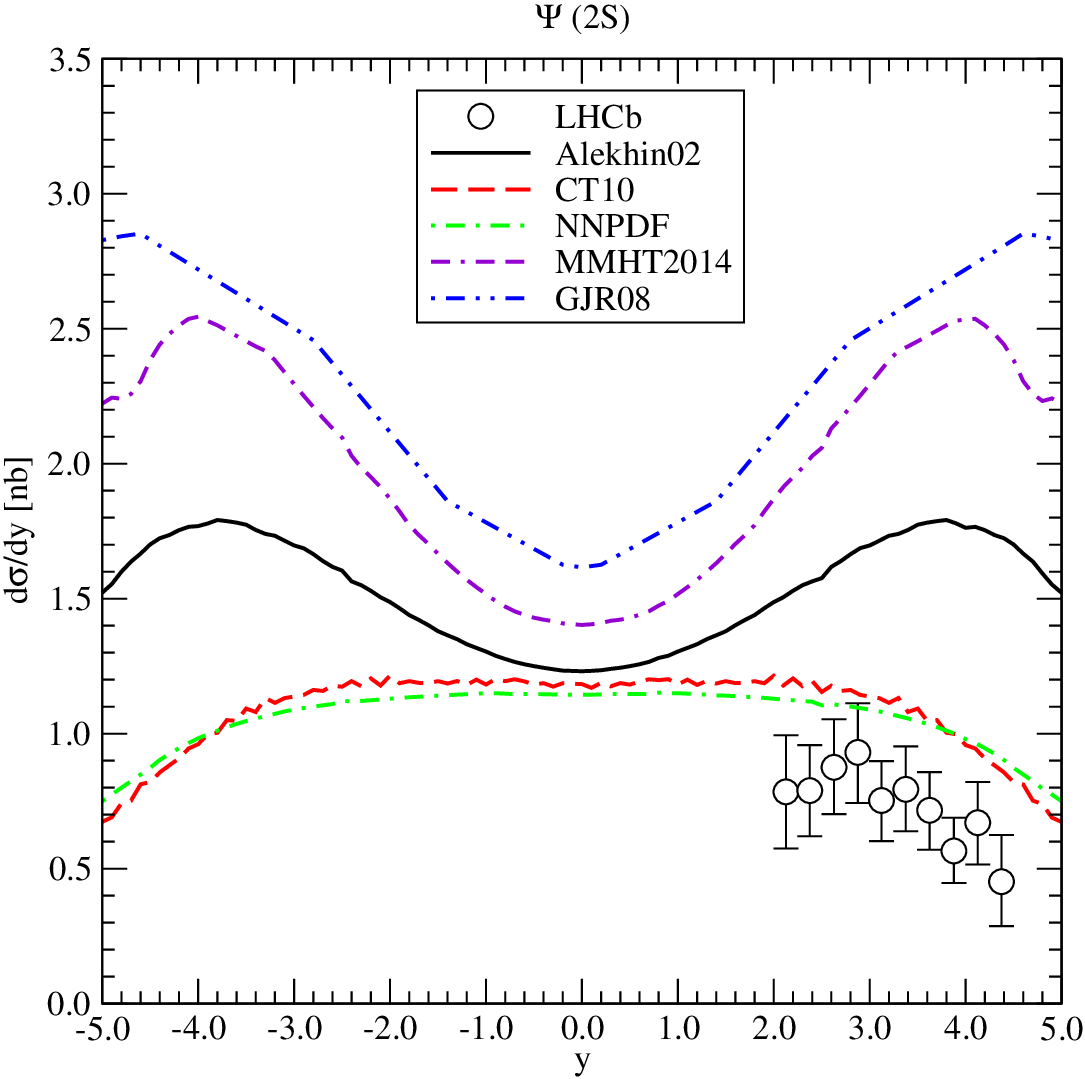}
\end{tabular}
\end{center}
\caption{\label{fig:psi2s} Left panel: Energy dependence of the exclusive  $\Psi(2S)$ photoproduction cross section obtained using the parameters described in the Table \ref{tab:param}. Right panel: Predictions for the rapidity distribution for the exclusive $\Psi(2S)$ photoproduction in $pp$ collisions at $\sqrt{s} = 7$ TeV. Data from  Refs.  
\cite{Adloff:2002re,lhcb2}.}
\end{figure}

\begin{figure}
\begin{center}
\begin{tabular}{cc}
 \includegraphics[scale=0.5]{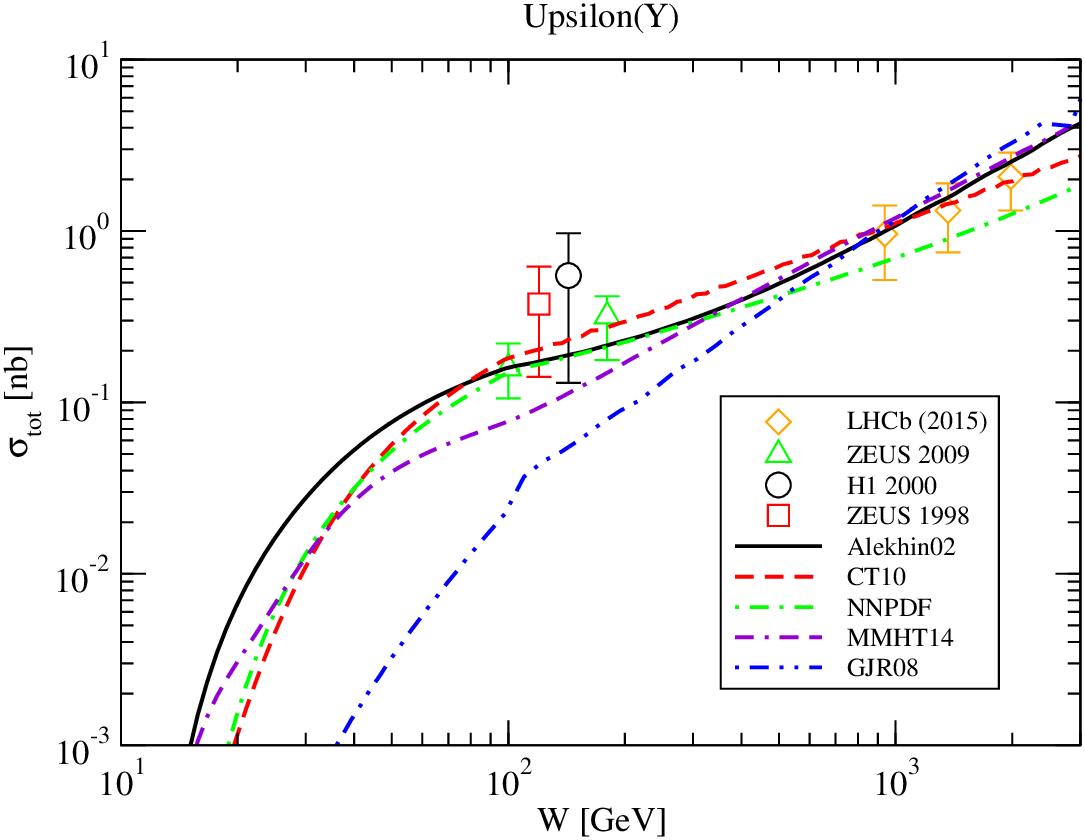} & \includegraphics[scale=0.38]{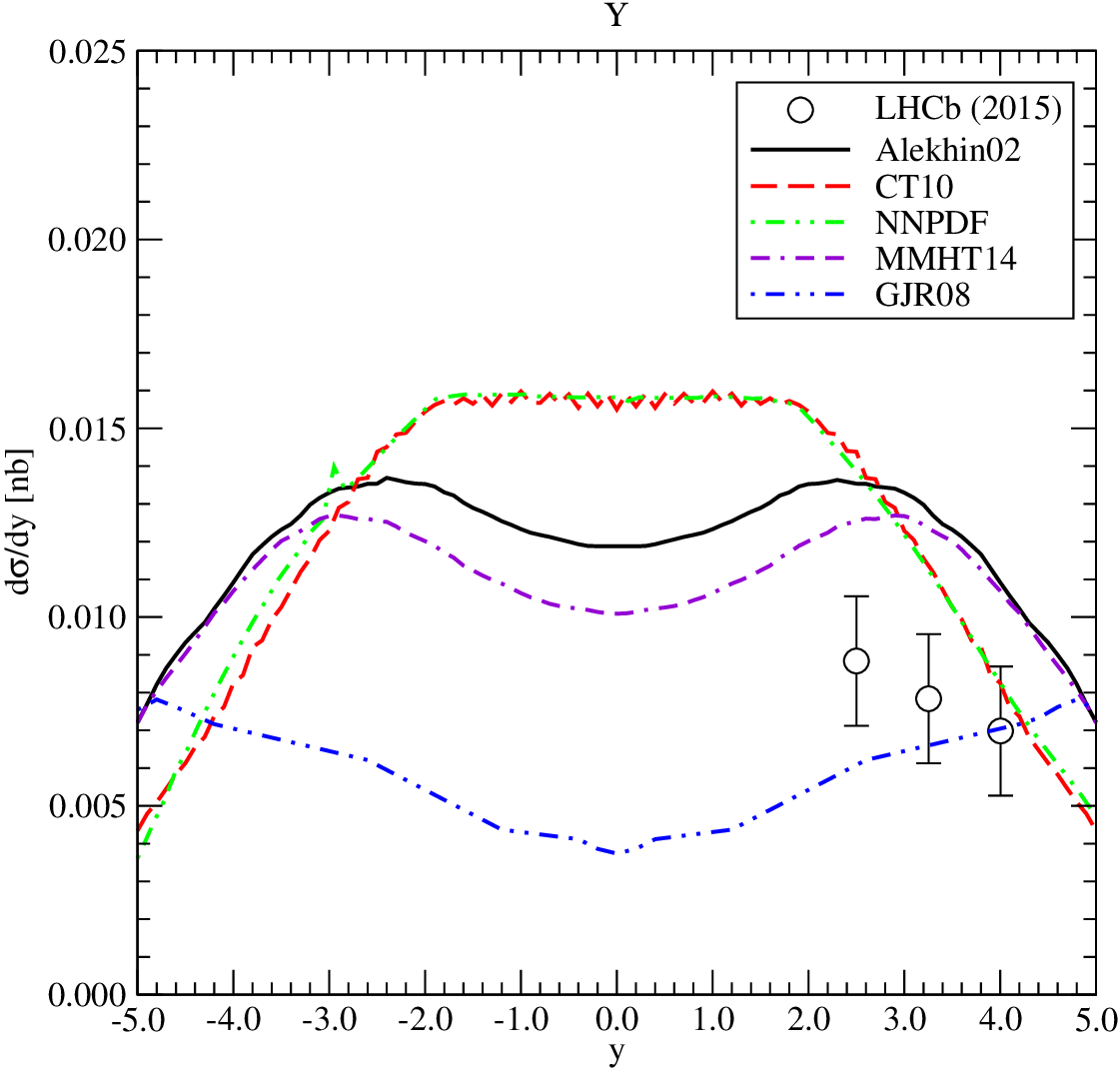}
\end{tabular}
\end{center}
\caption{\label{fig:upsilon} Left panel: Energy dependence of the exclusive  $\Upsilon$ photoproduction cross section obtained using the parameters described in the Table \ref{tab:param}. Right panel: Predictions for the rapidity distribution for the exclusive $\Upsilon$ photoproduction in $pp$ collisions at $\sqrt{s} = 7$ TeV. Data from  Refs.  
\cite{ZEUS_ups,H1_jpsi,Chekanov:2009zz,lhcb_ups}. }
\end{figure}

In Table \ref{tab:param} we present the fitted parameters for the different processes and a comparison between the predictions and the HERA data are shown in the left panel of the Figs. \ref{fig:jpsi}, \ref{fig:psi2s} and \ref{fig:upsilon}. In general, the results gives  reasonable values for the $\chi^2/\mathrm{d.o.f.}$,  inside the 98\% confidence level. We obtain smaller values of $\chi^2/\mathrm{d.o.f.}$ for heavier states,  mainly due the small number of experimental data. In contrast, for the $J/\psi$ data, a small $\chi^2/\mathrm{d.o.f.}$ is only obtained using  CT10 and NNPDF parametrizations, which describe the data in the full energy range, with the other parametrizations being able to describe the data in a restrict range. Moreover, as can be verified from the analysis of the Figs. \ref{fig:jpsi}, \ref{fig:psi2s} and \ref{fig:upsilon}, the predictions of the distinct parametrizations differ significantly in the kinematical range beyond the HERA data. It is important to emphasize that the results for  $\xi$ and ${\cal{N}}$ obtained using the CT10 and NNPDF parametrizations indicate that  
in order to describe the data for $J/\Psi$ and $\Psi(2S)$ production a large amount of higher order corrections are necessary. In contrast, these corrections are small for the $\Upsilon$ case.

In the right panel of the Figs. \ref{fig:jpsi}, \ref{fig:psi2s} and \ref{fig:upsilon} we present the corresponding predictions for the rapidity distributions for the vector meson photoproduction in $pp$ collisions at $\sqrt{s} = 7$ TeV. We compare our predictions with the recent data from the LHCb Collaboration \cite{lhcb,lhcb2,lhcb_ups}. As the rapidity distributions probe a large range of $\gamma p$ center of mass energies, the differences present in the predictions for  
$\sigma_{\gamma p \rightarrow V p}$ are amplified in $d\sigma/dy$, particularly for large rapidities in the case of the lighter mesons. For $J/\Psi$ and $\Psi(2S)$ production  the models characterized by a strong increasing of the gluon distribution at small - $x$ and small hard scales (Alekhin, MMHT and GJR) predict a double peak structure in the distributions. In contrast, the CT10 and NNPDF parametrizations predict a flat $y$ distribution in a large rapidity range. These parametrizations describe quite well the LHCb data for the $J/\Psi$ production and overestimate the $\Psi(2S)$ data. 
It is important to emphasize that we have obtained these predictions using the model for 
$\mathcal{S}^2(W_\pm)$ proposed in \cite{Jones1,Jones2}. Another aspect is the description of the 
$\Psi(2S)$ wave function used in our calculations, which is currently a subject of intense debate.
In the case of the $\Upsilon$ production, we obtain that the different models are not able to describe the current LHCb data, with the predictions at central rapidities being largely distinct. These results can indicate e.g.  smaller values for $\mathcal{S}^2(W_\pm)$ than those proposed in \cite{Jones1}. Certainly, the future CMS data for $y \approx 0$, which are currently under analysis, will be important to get more definitive conclusions.

\begin{table}[t]
\begin{center}
\begin{ruledtabular}
\begin{tabular}{c|ccc|ccc|ccc}
 & \multicolumn{3}{c|}{$J/\psi$} & \multicolumn{3}{c|}{$\Psi(2S)$} & \multicolumn{3}{c}{$\Upsilon$} \\ 
 Parametrization & $\xi$ & $\mathcal{N}$ & $\chi^2/\mathrm{d.o.f.}$ & $\xi$ & $\mathcal{N}$ & $\chi^2/\mathrm{d.o.f.}$ & $\xi$ & $\mathcal{N}$ & $\chi^2/\mathrm{d.o.f.}$ \\ \hline
 Alekhin02 & 0.879 & ~~0.180 & ~3.818 & 0.816 & ~~0.133 & 1.520 & 8.488$\cdot 10^{-2}$ & 1.087$\cdot 10^{-4}$ & 0.624 \\
 CT10      & 3.412 & ~45.041 & ~1.179 & 3.783 & ~59.331 & 1.682 & 0.614 & 0.188 & 0.312 \\
  NNPDF     & 4.373 & 139.670 & ~1.215 & 5.064 & 208.837 & 1.871 & 0.939 & 1.063 & 0.312 \\
 MMHT14    & 1.035 & ~~0.297 & ~7.226 & 0.641 & ~~0.101 & 1.220 & 0.135 & 3.690 & 1.820 \\
 GJR08     & 2.202 & ~~0.755 & 11.740 & 3.436 & ~~7.799 & 8.824 & 14.257 & 1.428$\cdot 10^{3}$ & 3.707
\end{tabular}
\end{ruledtabular}
\end{center}
\caption{\label{tab:param} Values of the free parameters $\xi$ and $\mathcal{N}$ for the different gluon parametrizations obtained by the minimization of $\chi^2/\mathrm{d.o.f.}$ for the distinct processes. The  $\chi^2/\mathrm{d.o.f.}$ values are show for comparison.}
\end{table}

A comment is in order. In our study we have considered leading order gluon distributions in order to be theoretically consistent with the fact that our expression for the total cross section have been derived at the leading logarithmic approximation. However, we have verified that our main conclusions are not modified if next - to - leading - order (NLO) gluon distributions are used as input in the calculations.  Basically, the NLO corrections  in the global analysis implies that the associated gluon distributions are not so steep at small - $x$ and low  values of the hard scale. As a consequence, a better description of the HERA data for the vector meson production is feasible. For example, in the case of the GJR parametrization, the value of $xg$ for $x = 10^{-4}$ and $\mu = M_{J/\Psi}/2$ decreases by a factor 2. The resulting values for the $\chi^2/\mathrm{d.o.f.}$ for the description of the HERA data becomes 
1.5 / 2.2 / 1.9 for the $J/\Psi / \Psi(2S) / \Upsilon$ production, respectively.  However, these values still are larger than those obtained using the LO and NLO CT10 parametrizations. Moreover, as the difference between the GJR and CT10 gluon distributions increases at smaller values of $x$, probed at larger values of $|y|$, the resulting GJR (NLO) predictions for the rapidity distributions are not able to describe the LHC data.  Similar results are obtained using the NLO gluon distributions derived in Refs. \cite{Alekhin:2002fv,Harland-Lang:2014zoa}. In the case of the CT10 and NNPDF parametrizations, we have that the impact of the NLO corrections is small, with the resulting gluon distributions being similar to those obtained at leading order. As a consequence, the associated predictions for the rapidity distributions are similar to those presented in the Figs.  \ref{fig:jpsi}, \ref{fig:psi2s} and \ref{fig:upsilon}.  These results indicate that the HERA and  LHC data are better described when the gluon distribution present a slow increasing at small - $x$ and low $Q$, in agreement with conclusion obtained at leading order.

Finally, lets summarize our main conclusions. Exclusive vector meson photoproduction offers a unique opportunity to constrain the gluon density of the proton in the kinematical range of small - $x$ and low $Q^2$, which is the range in which the global analysis do not reliable determine the gluon distribution. In this paper we have performed  a phenomenological study of this process using the leading logarithmic formalism and different models for $xg$ predicted by distinct groups that perform the global analysis of the experimental data in order to obtain the PDFs. We  demonstrated that the combination of anyone of these models and the leading logarithmic formalism  is not able to describe the experimental data for the $J/\Psi$ photoproduction. Assuming that the main modifications in the formalism due to higher order corrections are the change in the normalization of the cross section and in the value of the hard scale probed in the process, we have fitted the current experimental data for the $J/\Psi$, $\Psi(2S)$ and $\Upsilon$ photoproduction and used this new set parameters as input to calculate the rapidity distributions for the vector meson photoproduction in $pp$ collisions. We have demonstrated that a better description of the $\gamma h$ data  is obtained when the gluon distribution presents a slow increasing at small - $x$ and low $Q$. Moreover, our results indicated that the LHCb data also are better described by these models. However, results for $\Psi(2S)$ and $\Upsilon$ shown that the normalizations are not perfectly described, which can indicated   the a more detailed analysis about the magnitude of the factor $\mathcal{S}^2(W_\pm)$ for the different final states is important in future. Future experimental data, particularly at central rapidities, will be important to constrain the different models. 

A final comment is in order. In this paper, in order to estimate the exclusive photoproduction of vector mesons, we have assumed the leading logarithmic formalism and different solutions of the linear DGLAP equation. Our goal was to try to describe the current experimental data for the exclusive vector photoproduction by improving this formalism and assuming that non linear effects in the QCD dynamics can be disregarded in this process. Our results indicate that it is not an easy task. Certainly, more definitive conclusions could be obtained using the full NLO expression for the cross section, which still is in progress. However, it is important to emphasize that the vector meson photoproduction at HERA and LHC is quite well described using the color dipole formalism, taking into account the non linear effects in the QCD dynamics and assuming $\mathcal{S}^2(W_\pm) = 1$ (See e.g. Refs \cite{bruno1,bruno2}). Such aspects demonstrated that future  experimental results for the exclusive vector meson photoproduction in hadronic colliders are fundamental to understanding the QCD dynamics at high energies.

\begin{acknowledgements}
This work was  partially financed by the Brazilian funding
agencies CNPq, CAPES and FAPERGS.
\end{acknowledgements}

\end{document}